\documentclass{JHEP3}
\usepackage{psfrag,epsf,amsmath,amssymb,graphicx,cite}
\usepackage{scalefnt}
\newcommand{\note}[1]{}

\newcommand{\abbrev}{\scalefont{.9}}
\newcommand{\dred}{{\abbrev DRED}}
\newcommand{\dreg}{{\abbrev DREG}}
\newcommand{\ep}{\epsilon}
\newcommand{\api}{\frac{\alpha_s}{\pi}}
\newcommand{\eqn}[1]{Eq.\,(\ref{#1})}
\newcommand{\fig}[1]{Fig.\,\ref{#1}}

\newcommand{\sct}[1]{Sect.\,\ref{#1}}
\newcommand{\dd}{{\rm d}}

\newcommand{\order}[1]{{\cal O}(#1)}
\newcommand{\lo}{{\abbrev LO}}
\newcommand{\nlo}{{\abbrev NLO}}
\newcommand{\nnlo}{{\abbrev NNLO}}
\newcommand{\msbar}{\mbox{$\overline{\mbox{\abbrev MS}}$}}
\newcommand{\bld}[1]{\boldmath{$#1$}}

\newcommand{\oddhiggs}{A}

\newcommand{\topo}{\tilde {\cal O}}

\newcommand{\vep}{\varepsilon}
\newcommand{\qcd}{{\abbrev QCD}}
\newcommand{\sm}{{\abbrev SM}}
\newcommand{\mssm}{{\abbrev MSSM}}

\renewcommand{\Re}{{\rm Re}}

\newcommand{\mstop}[1]{m_{\tilde t_{#1}}}
\newcommand{\mgluino}{m_{\tilde g}}
\newcommand{\susy}{{\abbrev SUSY}}
\newcommand{\muSUSY}{\mu_{\rm SUSY}}
\newlength{\figwid}
\newcommand{\feynsl}[1]{
  \setbox0=\hbox{/} \setbox1=\hbox{$#1$}
  \dimen0=\wd0 \advance\dimen0 by -\wd1 \divide\dimen0 by 2
  \ifdim\wd0>\wd1 \lower.15ex
          \copy0\kern-\wd0\kern\dimen0\copy1\kern\dimen0
  \else \kern-\dimen0\lower.15ex
          \copy0\kern-\dimen0\kern-\wd1\copy1\fi}

\def\readRCS$#1: #2,v #3 #4 #5${%
 \def\filename{#2}%
 \def\fileversion{#3}%
 \def\filedate{#4}%
}

\readRCS
 $Id: ggasusy.tex,v 1.10 2005/11/01 11:41:23 rharland Exp $

\title{Pseudo-scalar Higgs production at next-to-leading order SUSY-QCD}
\author{Robert V. Harlander and Franziska Hofmann\\
  {\it Institut f\"ur Theoretische Teilchenphysik,
  Universit\"at Karlsruhe\\
  D-76128 Karlsruhe, Germany}\\
    E-mail: \email{robert.harlander@cern.ch},
\email{fhofmann@particle.uni-karlsruhe.de}\\
}

\preprint{\hepph{0507041}, \sf TTP05--07, SFB/CPP-05-17 --- July 2005}

\abstract{ The production rate of the {\abbrev CP}-odd Higgs boson in
the Minimal Supersymmetric Standard Model is evaluated through
next-to-leading order in the strong coupling constant.  The divergent
integrals are regulated using Dimensional Reduction, with a
straightforward implementation of $\gamma_5$. The result is confirmed
within Dimensional Regularization where $\gamma_5$ is implemented
according to the Standard Model calculation of Chetyrkin {\it et
al.}~\cite{Chetyrkin:1998mw}.  The well-known Standard Model result is
recovered if the masses of the supersymmetric particles tend to
infinity. }

\begin{document}
\setlength{\figwid}{7em}

\section{Introduction}\label{sec::Intro}
The Minimal Supersymmetric Standard Model (\mssm{}) predicts a
fundamental {\abbrev CP}-odd scalar particle $A$, commonly referred to
as the pseudo-scalar Higgs boson. One of the most important properties
that distinguishes it from its {\abbrev CP}-even analogues $h$ and $H$
is the absence of tree-level couplings to the electro-weak gauge bosons
$W$ and $Z$. Decay and production processes through these particles,
which have been shown to be extremely helpful for {\abbrev CP}-even
Higgs searches and studies, are thus very much suppressed.  This leaves
associated $t\bar tA$ and $b\bar bA$ production as well as the
loop-induced gluon fusion process as the most important production modes
of a pseudo-scalar Higgs boson at the Large Hadron Collider ({\abbrev
LHC}) (for a recent review, see Ref.\,\cite{Djouadi:2005gij}).  In this
paper, we present the evaluation of the inclusive gluon fusion cross
section through next-to-leading order (\nlo{}) in the strong coupling
constant.

Despite the fact that the tree-level $gg\phi$ coupling vanishes
($\phi\in\{h,H,A\}$), gluon fusion in general has a comparatively large
cross section because of the high gluon luminosity at the {\abbrev LHC}, and
the large top-Yukawa coupling. In the limit where squarks and gluinos
are decoupled, {\abbrev QCD} corrections have been evaluated through
\nnlo{}~\cite{Harlander:2002wh,Harlander:2002vv,Anastasiou:2002wq,
Anastasiou:2002yz,Ravindran:2003um} by employing an effective Lagrangian
for heavy top quarks (cf.\,\sct{sec::eff}); they have been shown to be
numerically significant but perturbatively well-behaved.

A precise determination of the gluon fusion cross section at the
{\abbrev LHC} could yield sensitivity to as yet undiscovered particles
that may mediate the $gg\phi$ coupling apart from the top and bottom
quarks. Within the \mssm{}, for example, top squarks can play a
significant role if they are lighter than around 400\,GeV (recall that
the Yukawa coupling for squarks is typically proportional to $m_q^2$
rather than $m_{\tilde q}^2$). In the case of {\abbrev CP}-even Higgs
bosons, such effects occur already at leading order (\lo{}) (i.e.,
1-loop). The \nlo{} result was obtained in Ref.\,\cite{Harlander:2004tp}
within an effective theory for top, stop, and gluino masses much larger
than the Higgs mass.

For the {\abbrev CP}-odd Higgs boson, however, squarks do not affect the
$ggA$ vertex at 1-loop level due to the structure of the $A\tilde
q\tilde q$ coupling, as will be shown below. The two-loop effects are
thus expected to have a larger influence than for the {\abbrev CP}-even
Higgs production. What adds to this is that, in the limit of large
$m_q$, the quark mediated contribution to the $ggA$ vertex does not
receive any \qcd{} corrections owing to the Adler-Bardeen theorem. As we
will show, the only 2-loop \qcd{} effects to this coupling are due to
mixed gluino-quark-squark diagrams in this limit, leading to a
potentially increased sensitivity to the gluino mass. Of course, the
heavy quark limit is not applicable for the bottom mediated gluon-Higgs
coupling which does receive \qcd{}
corrections~\cite{Spira:1993bb,Spira:1995rr,Spira:1997dg}.

The calculation involves a technical issue that deserves special
mention, namely the implementation of $\gamma_5$ within Dimensional
Reduction (\dred{}).  Its
mathematically consistent and practically feasible formulation has been
a subject of interest for many years now~\cite{Siegel:1980qs} (for
recent developments concerning \dred{}, see
Ref.\,\cite{Smith:2004ck,Stockinger:2005gx}).  Here we adopt an approach
close to the prescription of
Refs.\,\cite{'tHooft:fi,Breitenlohner:1977hr}. In addition, we perform
the calculation using Dimensional Regularization (\dreg{}) with
$\gamma_5$ implemented as in Ref.\,\cite{Chetyrkin:1998mw} and find full
agreement with the result obtained in \dred{}.

Another important consistency check for our calculation is obtained from
the \sm{} limit: even for an infinitely heavy supersymmetric (\susy{})
spectrum, the \susy{} diagrams give a non-vanishing contribution which
exactly cancels the top mass counter-term contribution arising from the
\lo{} \sm{} diagram.  The well-known result of vanishing higher order
corrections to the quark-mediated $ggA$ coupling, as required by the
Adler-Bardeen theorem, is thus recovered in a non-trivial way.
Combining this observation with the considerations of
Ref.\,\cite{Stockinger:2005gx}, it seems possible that in a
supersymmetric theory the formal difficulties of \dred{} do not pose
serious technical problems in practical calculations.

The outline of this paper is as follows: In \sct{sec::notation} we
introduce our notation and quote the \lo{} result for the partonic
process $gg\to A$. In \sct{sec::higher}, the effective Lagrangian
underlying our calculation is introduced and the treatment of $\gamma_5$
is discussed. \sct{sec::results} outlines the method of the calculation
and discusses the general structure of the result. It also provides
analytic formulae in some limiting cases. The numerical influence of the
\nlo{} terms is discussed in \sct{sec::discussion}. In
\sct{sec::conclusions} our findings are summarized and an outlook on
possible extensions of this work is given.

\section{Notation and Leading Order Result}\label{sec::notation}

\subsection{Lagrangian}
We write the underlying Lagrangian in the following form:
\begin{equation}
\begin{split}
{\cal L} &= {\cal L}_{\rm QCD} + {\cal L}_{ qA}
+ {\cal L}_{\rm SQCD} + {\cal L}_{\tilde qA}\,,
\label{eq::lag}
\end{split}
\end{equation}
where
\begin{equation}
\begin{split}
{\cal L}_{qA} &= i\sum_{q} \frac{m_q}{v}\,g_q^A\, \bar q\gamma_5 q
A\,,\qquad {\cal L}_{\tilde qA} = \sum_{q}\sum_{i=1}^{2}
\frac{m_q^2}{v}\,\tilde g_{q,ij}^A\, \tilde q_i\tilde q_j\,A\,.
\label{eq::lags}
\end{split}
\end{equation}
The sums $\sum_{q}$ run over all quark flavors.  $m_q$ is the mass of
the quark $q$ and $v\approx 246$\,GeV the Higgs vacuum expectation value;
the coupling constants $g_q^A$ and $\tilde g_{q,ij}^A$ will be defined
below.  ${\cal L}_{\rm QCD}$ denotes the full {\abbrev QCD} Lagrangian
with six quark flavors, while ${\cal L}_{\rm QCD}+{\cal L}_{\rm SQCD}$
is the supersymmetric extension of ${\cal L}_{\rm QCD}$ within the
\mssm{}, i.e., ${\cal L}_{\rm SQCD}$ incorporates kinetic, mass, mixing,
and interaction terms of all the squarks and gluinos.  Since we will be
concerned with higher orders in the strong coupling $\alpha_s$ only, $A$
does not appear as a dynamical field and does not require a kinetic
or a mass term.

$\tilde q_1, \tilde q_2$ denote the squark mass eigenstates which are
related to
the chiral eigenstates $\tilde q_L, \tilde q_R$ (the superpartners of
the left- and the right-handed quark $q$) through\note{check}
\begin{equation}
\begin{split}
\left(
\begin{array}{c}
{\tilde q_1} \\
{\tilde q_2}
\end{array}
\right)
 &=
\left(
\begin{array}{cc}
\cos\theta_q & \sin\theta_q\\
-\sin\theta_q & \cos\theta_q
\end{array}
\right)
\left(
\begin{array}{c}
{\tilde q_L} \\
{\tilde q_R}
\end{array}
\right)\,.
\end{split}
\end{equation}
The coupling constants relevant for the following discussion
are\footnote{For a more complete collection of Feynman rules, see
Ref.\,\cite{Kraml-Eberl}, for example.}
\begin{equation}
\begin{split}
g_b^A &= \tan\beta\,,\qquad g_t^A = \cot\beta\,,\qquad
\tilde g_{t,11}^A = \tilde g_{t,22}^A = 0\,,\\
\tilde g_{t,12}^A &= - \tilde g_{t,21}^A = 
\frac{\mstop{1}^2 - \mstop{2}^2}{2m_t^2}\,\sin 2\theta_t\,\cot\beta
+ \frac{\muSUSY}{m_t}\,(\cot^2\beta +1 )\,.
\label{eq::couplings}
\end{split}
\end{equation}
The effect of quarks other than bottom and top can be neglected because of
their small Yukawa couplings. Bottom and sbottom effects are typically
of order $(m_b^2/M_A^2)\tan^2\beta$ relative to the top and stop
effects. In this first analysis of \nlo{} squark effects to
pseudo-scalar Higgs production, we neglect the effect of sbottom quarks;
they require a further extension of our approach.  Our results are
thus strictly valid only for not-too-large values of $\tan\beta$.  Pure
bottom effects, on the other hand, will be included through \nlo{} in
our numerical analysis.

\subsection{Hadronic cross section}
The cross section for the hadronic process $pp\to A+X$ at a
center-of-mass energy $\sqrt{s}$ is determined by the formula\footnote{The
  modifications for $p\bar p$ collisions are obvious and shall not be
  pointed out explicitely in this paper.}
\begin{equation}
\begin{split}
\sigma(z) &= \sum_{i,j\in\{q,\bar q,g\}}
\int_z^1\dd x_1\int_{z/x_1}^1\dd
x_2\,\varphi_i(x_1)\varphi_j(x_2)\,\hat\sigma_{ij}
\left(\frac{z}{x_1x_2}\right)\,,
\qquad z:= \frac{M_A^2}{s}\,,
\label{eq::hadpart}
\end{split}
\end{equation}
where $\varphi_i(x)$ is the density of parton $i$ inside the proton.
$\hat\sigma_{ij}$ is the cross section for the process $ij\to A+X$,
where, as indicated in \eqn{eq::hadpart}, $i$ and $j$ are parton labels.
For our numerical analysis, we will use the {\abbrev MRST} parton
density sets throughout this paper~\cite{Martin:2001es,Martin:2003sk}.

Sample diagrams for $ij\!\!=\!\!gg, gq, q\bar q$ that contribute to the
inclusive Higgs production rate at \lo{} and \nlo{} are shown in
\fig{fig::lonlo}.

\FIGURE{
    \begin{tabular}{c}
      \begin{tabular}{cc}
      \includegraphics[bb=154 450 450
      660,width=8em]{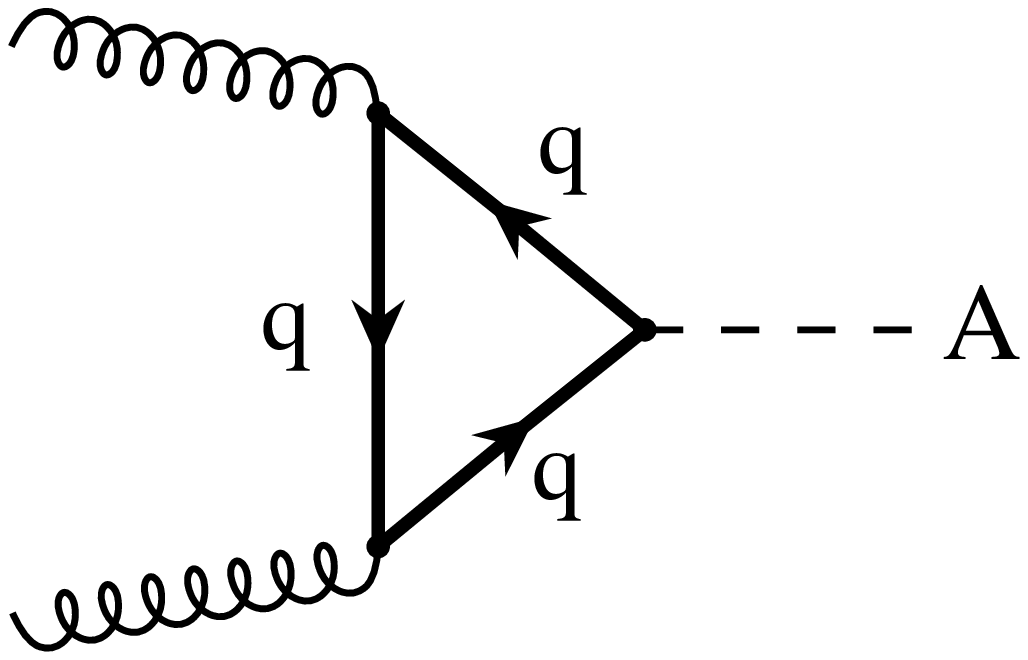} &
      \includegraphics[bb=154 450 450
      660,width=8em]{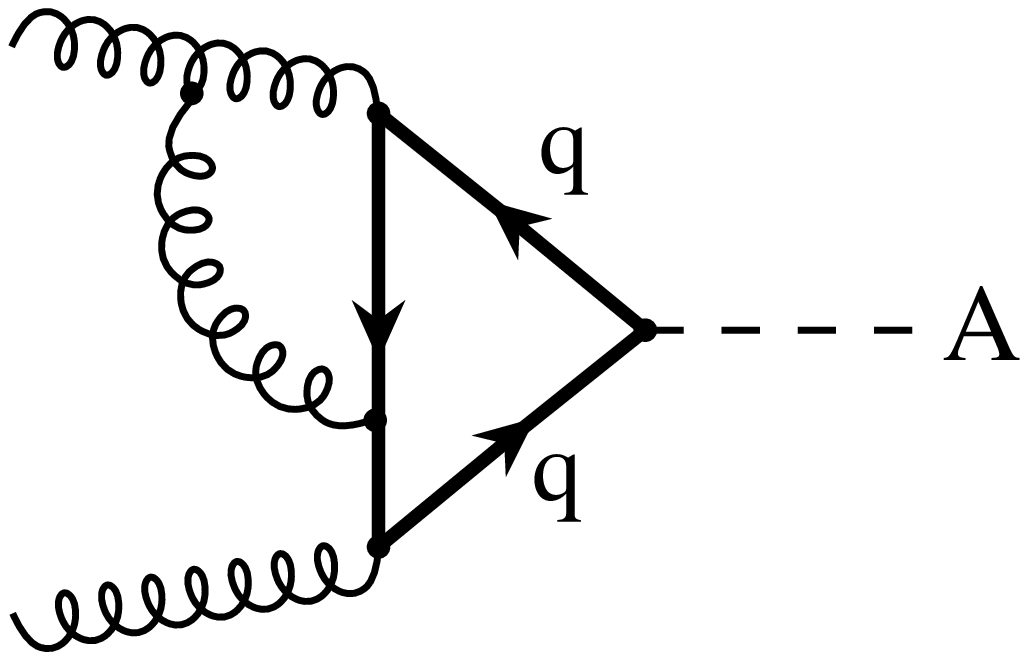} \\[-.5em]
      (a) & (b)
      \end{tabular}
      \\[4.5em]
      \begin{tabular}{cccc}
      \includegraphics[bb=154 450 450
      660,width=8em]{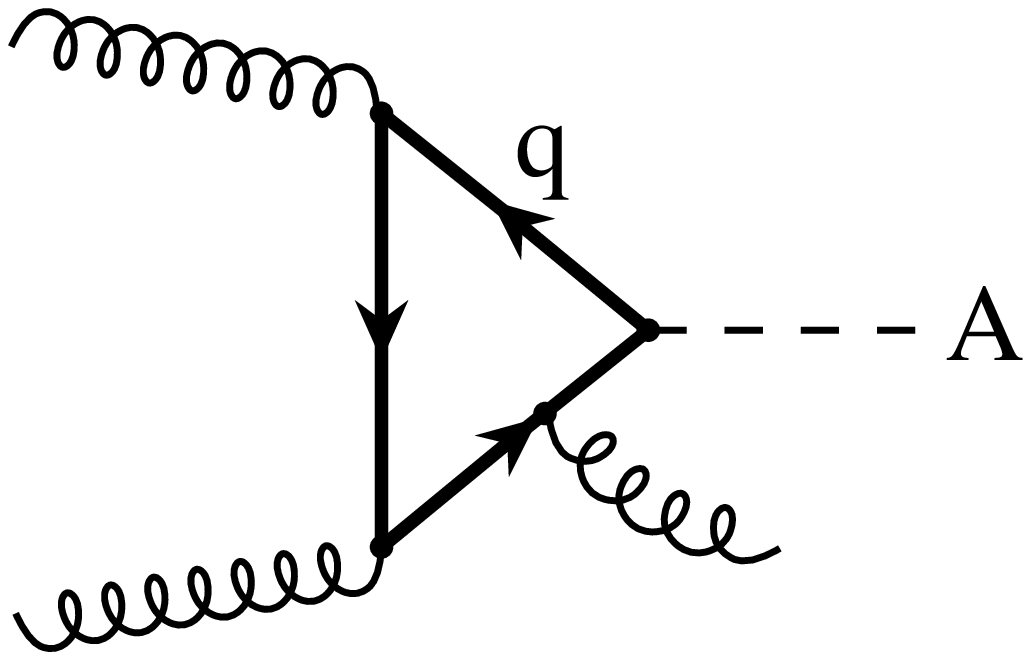} &
      \includegraphics[bb=154 450 450
      660,width=7.5em]{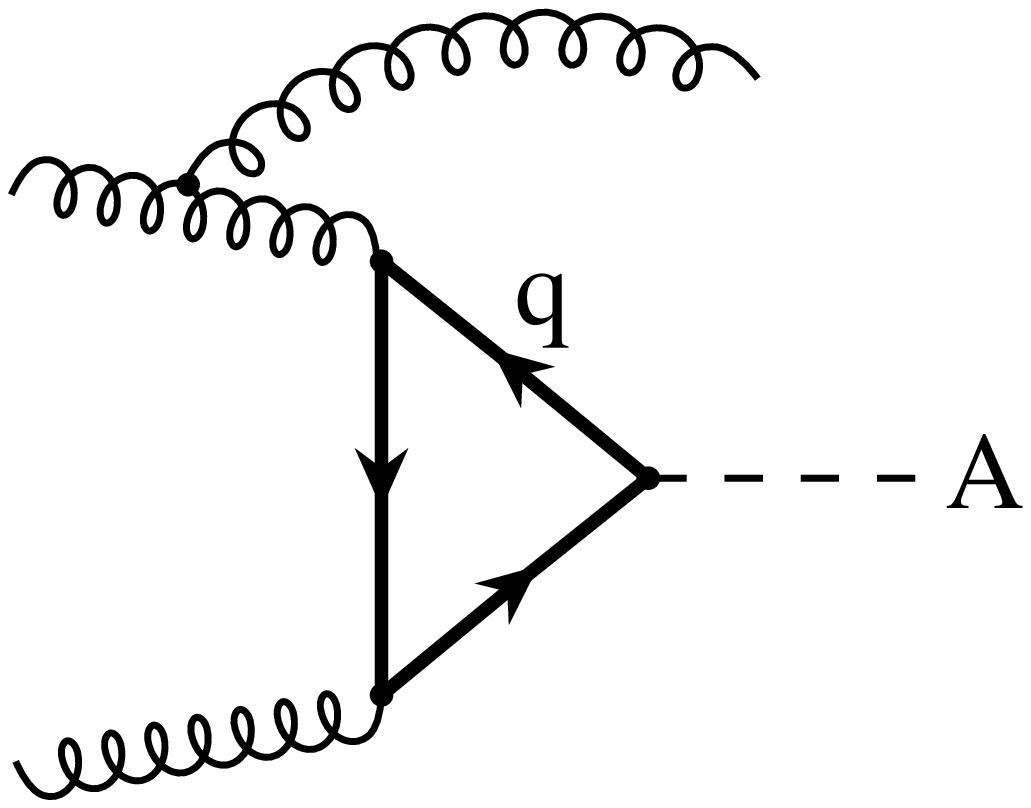} &
      \includegraphics[bb=154 450 450
      660,width=7em]{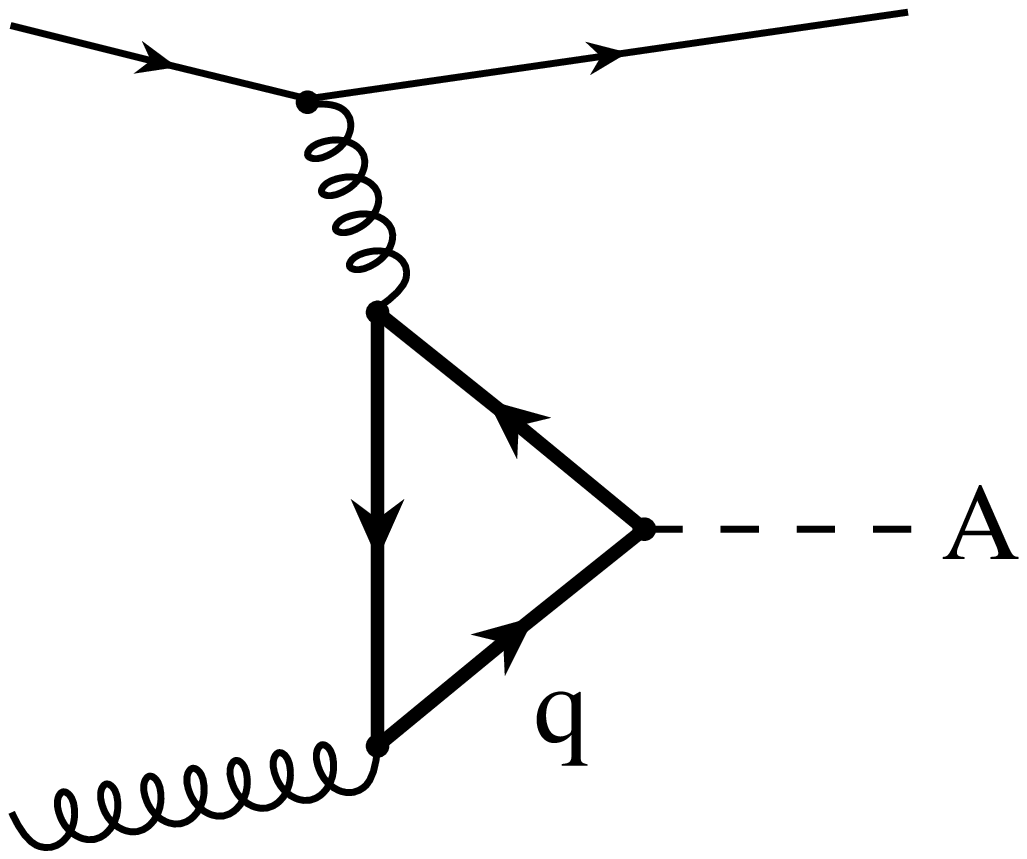} &
      \includegraphics[bb=100 500 410
      680,width=9em]{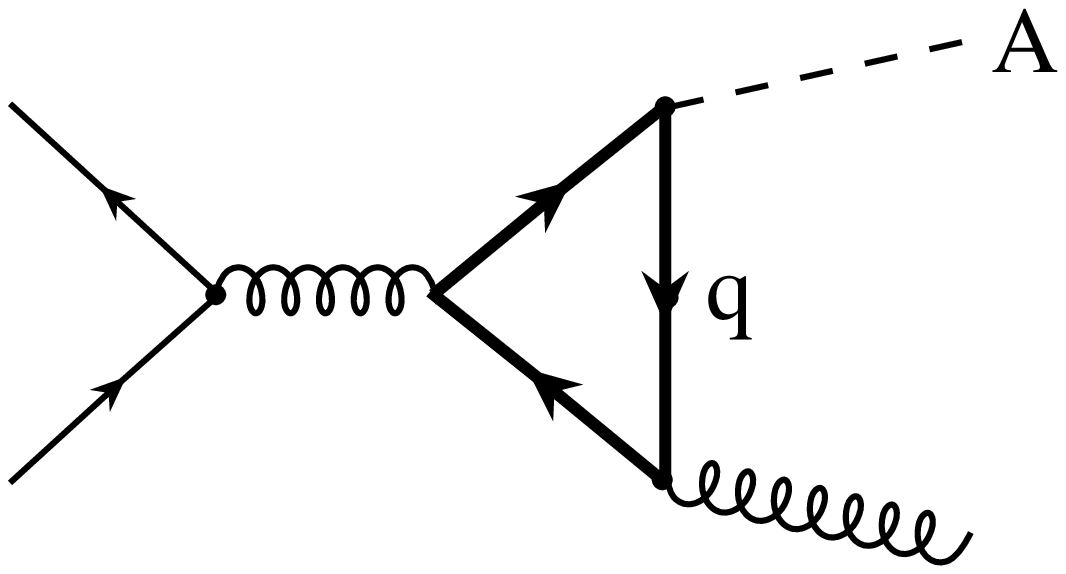} \\[-.5em]
      (c) & (d) & (e) & (f)
      \end{tabular}
      \\[4.5em]
      \begin{tabular}{cccc}
      \includegraphics[bb=154 450 450
      660,width=8em]{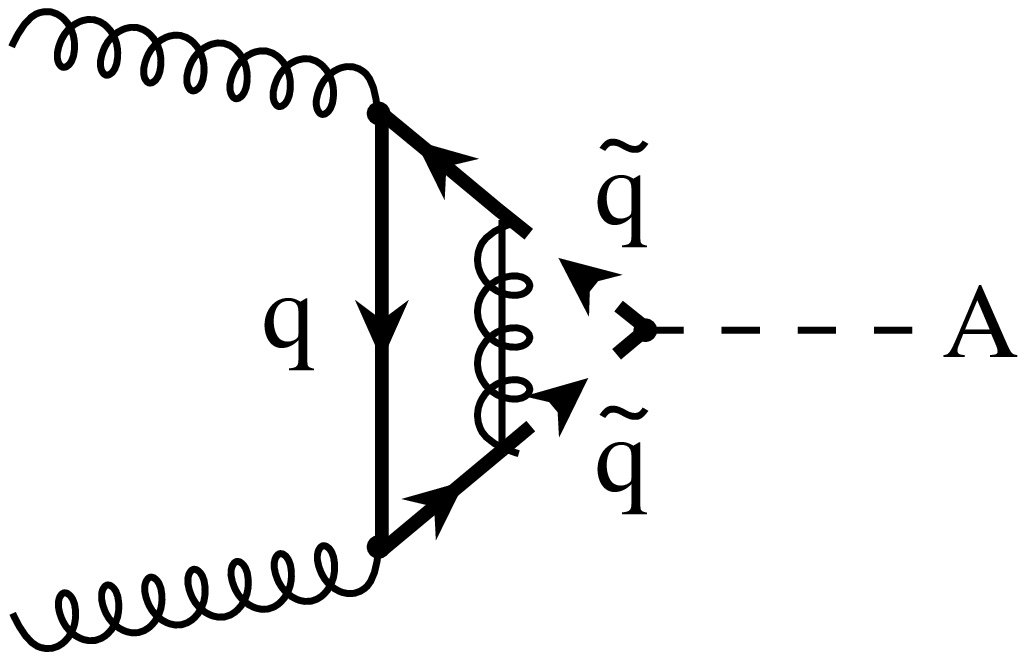} &
      \includegraphics[bb=154 450 450
      660,width=8em]{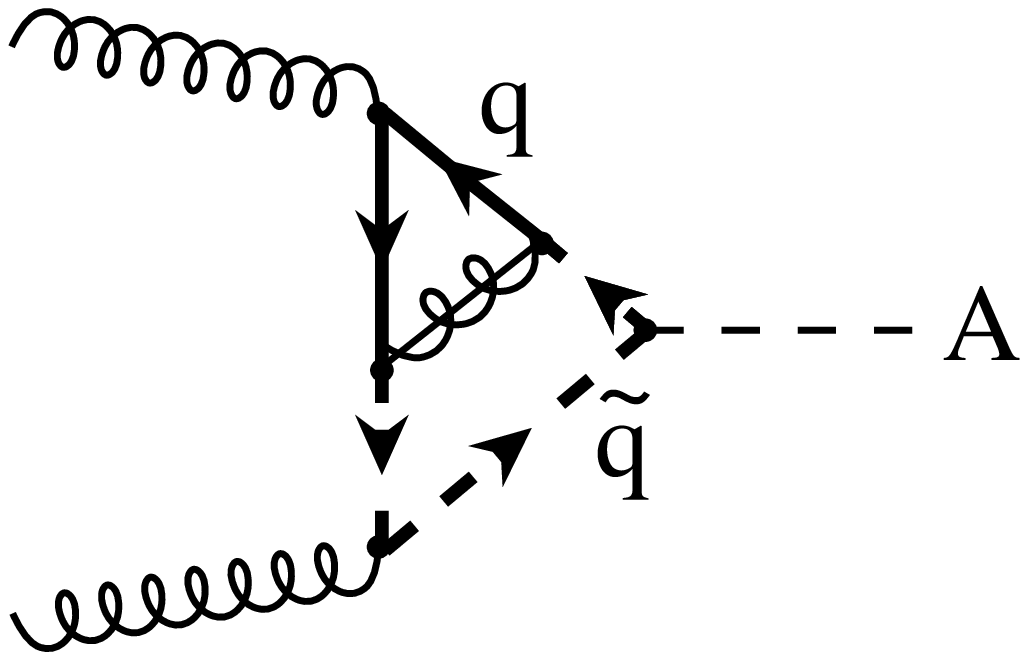} &
      \includegraphics[bb=154 450 450
      660,width=8em]{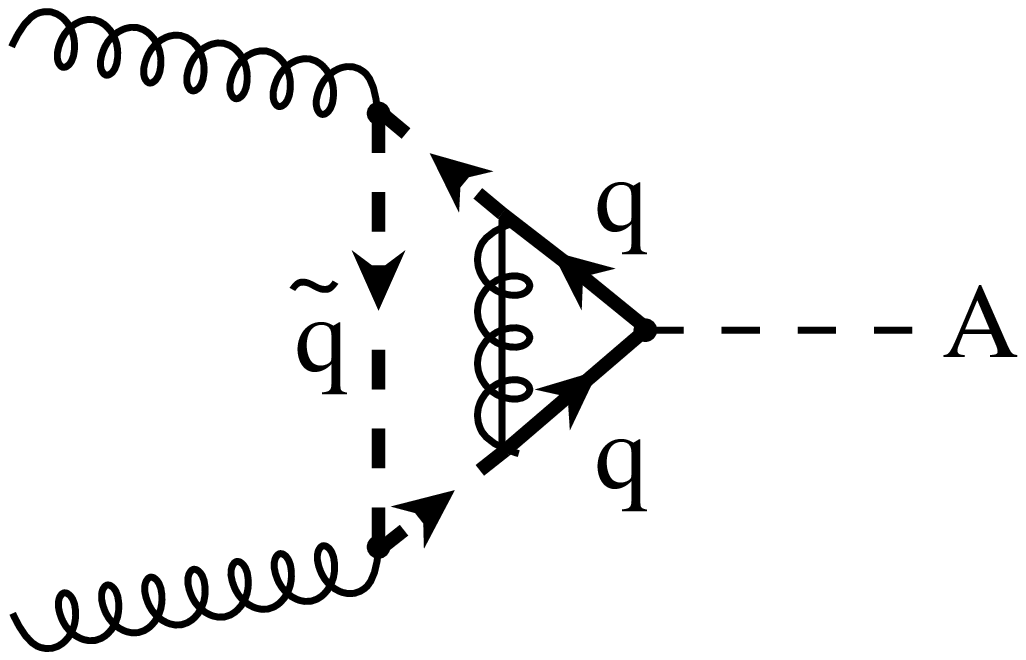} &
      \includegraphics[bb=154 450 450
      660,width=8em]{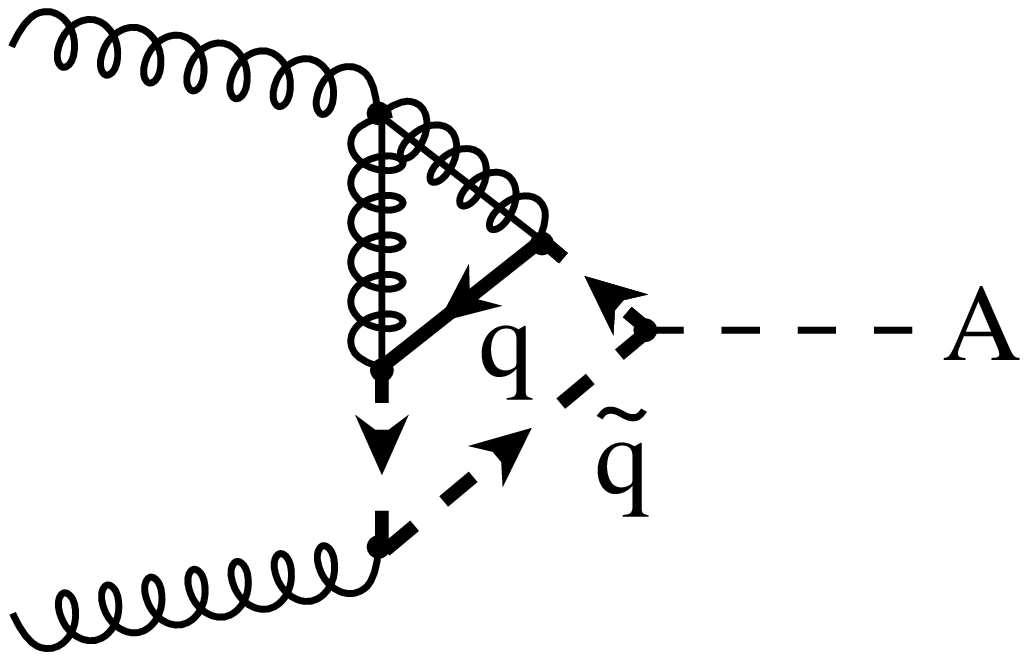}\\[-.5em]
      (g) & (h) & (i) & (j)
      \end{tabular}
    \end{tabular}
    \caption[]{\sloppy
      Diagrams contributing to the process $pp\to A+X$ in the
      \mssm{} at (a) \lo{} and (b)--(j) \nlo{}.
      According to the definition of the main text, (a)--(f) are referred
      to as ``\sm{} diagrams'', (g)--(j) as ``\susy{} diagrams''.
      The straight-curly line connecting quarks and squarks represents a
      gluino (denoted by $\tilde g$ in the text).
      \label{fig::lonlo}%
    }
}

\subsection{Standard Model limit and leading order result}
In the limit where all {\abbrev SUSY} masses tend to infinity (denoted
in the following by $M_{\rm SUSY}\to \infty$), the Lagrangian of
\eqn{eq::lag} reduces to
\begin{equation}
\begin{split}
{\cal L}^{\rm SM} &= {\cal L}_{\rm QCD} + {\cal L}_{qA}\,.
\label{eq::lagsm}
\end{split}
\end{equation}
This will be called the ``Standard Model (\sm{})'' limit in what follows,
despite the fact that the \sm{} does not contain a {\abbrev CP}-odd
Higgs boson. Consequently, diagrams without any squark and gluino lines
will be called ``\sm{} diagrams'' (e.g. \fig{fig::lonlo}\,(a)--(f)),
while ``\susy{} diagrams'' contain at least one propagator of a \susy{}
particle (\fig{fig::lonlo}\,(g)--(j)).

Note that due to the antisymmetric structure of the $\tilde g_{q,ij}^A$,
see \eqn{eq::couplings}, there are no \susy{} diagrams at the one-loop
level. The \lo{} result for the partonic process $gg\to A$ is thus
determined solely by diagrams of the type shown in
\fig{fig::lonlo}\,(a). It reads
explicitely~\cite{Kalyniak:1985ct,Bates:1986zv,Gunion:1988mf,hunter}:
\begin{equation}
\begin{split}
\hat\sigma_{gg}(x) &= \sigma_{0}^A\,\delta(1-x) +
\order{\alpha_s^3}\,,\\
\sigma_{0}^A &= \frac{\pi}{256 v^2}\left(\api\right)^2\,
\bigg|\sum_{q\in\{t,b\}} {\cal A}_q(\tau_q)\bigg|^2\,,\\
{\cal
A}_q(\tau_q) &= g_q^A\, \tau_q f(\tau_q)\,,\qquad\tau_q =
\frac{4m_q^2}{M_A^2}\,,
\label{eq::LO}
\end{split}
\end{equation}
where
\begin{equation}
  \begin{split}
    f(\tau) = \left\{
    \begin{array}{ll}
      \displaystyle\arcsin^2\frac{1}{\sqrt{\tau}}\,, & \tau \geq 1\,,\\
      \displaystyle -\frac{1}{4}\left[
	\ln\frac{1+\sqrt{1-\tau}}{1-\sqrt{1-\tau}} - i\pi\right]^2\,, &
      \tau < 1\,.
    \end{array}
    \right.
    \label{eq::ftau}
  \end{split}
\end{equation}
Throughout this paper, $\alpha_s$ denotes the strong coupling constant,
renormalized in the $\msbar{}$ scheme for \qcd{} with five massless
quark flavors.

\section{Higher orders}\label{sec::higher}

\subsection{Standard Model result -- Effective Lagrangian}\label{sec::eff}
The \sm{} contributions, based on ${\cal L}_{\rm SM}$ of
\eqn{eq::lagsm}, are known through \nlo{} in terms of 1-dimensional
integral representations~\cite{Spira:1993bb,Spira:1995rr}, implemented
in the program {\tt HIGLU}~\cite{Spira:1995mt}. They include both top
and bottom quark loops as shown in \fig{fig::lonlo}\,(a)--(f)
($q\in\{b,t\}$).

It has been shown in Ref.\,\cite{Spira:1995rr,Spira:1997dg} that the
\nlo{} top quark contributions are well approximated by the formula
\begin{equation}
\begin{split}
\sigma_t^{\rm NLO} &= K_\infty\,\sigma_t^{\rm LO}\,,\quad
\mbox{where}\qquad
K_\infty = \frac{\sigma_\infty^{\rm NLO}}{\sigma_\infty^{\rm LO}}\,.
\end{split}
\end{equation}
$\sigma_{\infty}$ is the cross section evaluated in an
effective theory, obtained by integrating out the top quark:
\begin{equation}
\begin{split}
{\cal L}_{\rm QCD} + {\cal L}_{qA} \stackrel{m_t\gg M_A}{\longrightarrow}
{\cal L}_{ggA}^{\rm SM} = 
    -\frac{\oddhiggs{}}{v}\left[\tilde C_1^{\rm SM}\, \topo_1 +
    \tilde C_{2}^{\rm SM}\, \topo_{2}\right]
    + {\cal L}_{\rm QCD}^{(5)}\,.
    \label{eq::leffsm}
\end{split}
\end{equation}
The operators are defined as
\begin{equation}
\begin{split}
 \topo_1 &= \frac{1}{4}
    \vep^{\mu\nu\alpha\beta}G^{a}_{\mu\nu} G^{a}_{\alpha\beta}\,,\qquad
    \topo_{2} = \sum_{q\neq t}\partial_\mu\left( \bar q
    \gamma^\mu\,\gamma_5 q\right)\,.  \label{eq::leff}
\end{split}
\end{equation}
Here, $\{q\} := \{d,u,s,c,b\}$ denotes the set of light (in our case
massless) quark fields and $G_{\mu\nu}^a$ the gluon field strength
tensor. ${\cal L}_{\rm QCD}^{(5)}$ is the Standard {\abbrev QCD}
Lagrangian with five quark flavors.

The Wilson coefficients $\tilde C_1^{\rm SM}$ and $\tilde C_{2}^{\rm
SM}$ have been calculated through \nnlo{} in
Ref.\,\cite{Chetyrkin:1998mw}. $\tilde C_2^{\rm SM}$ contributes only at
\nnlo{} and shall not be discussed any further in this paper. The result
for $\tilde C_1^{\rm SM}$ is
\begin{equation}
\begin{split}
\tilde C_1^{\rm SM} &= -\cot\beta\frac{\alpha_s}{16\pi} +
\order{\alpha_s^4}\,,
\label{eq::c1sm}
\end{split}
\end{equation}
as it had been previously suggested in
Refs.\,\cite{Kauffman:1993nv,Djouadi:1993ji,Spira:1993bb} on the basis
of the Adler-Bardeen theorem~\cite{adlerbardeen}.

$\topo_1$ generates vertices which couple two and three gluons to the
pseudo-scalar Higgs boson (the $ggggA$-vertex vanishes according to
the Jacobi identity of the structure functions of SU(3)).  Sample
diagrams that contribute to the \nlo{} cross section in the effective
theory of \eqn{eq::leffsm} are shown in \fig{fig::process}. The full set
has been calculated through \nlo{} in
Ref.\,\cite{Spira:1993bb,Spira:1995rr}, and through \nnlo{} in
Ref.\,\cite{Harlander:2002vv,Anastasiou:2002wq,Ravindran:2003um}.

\FIGURE{%
    \begin{tabular}{c}
      \begin{tabular}{ccc}
      \includegraphics[width=8em]{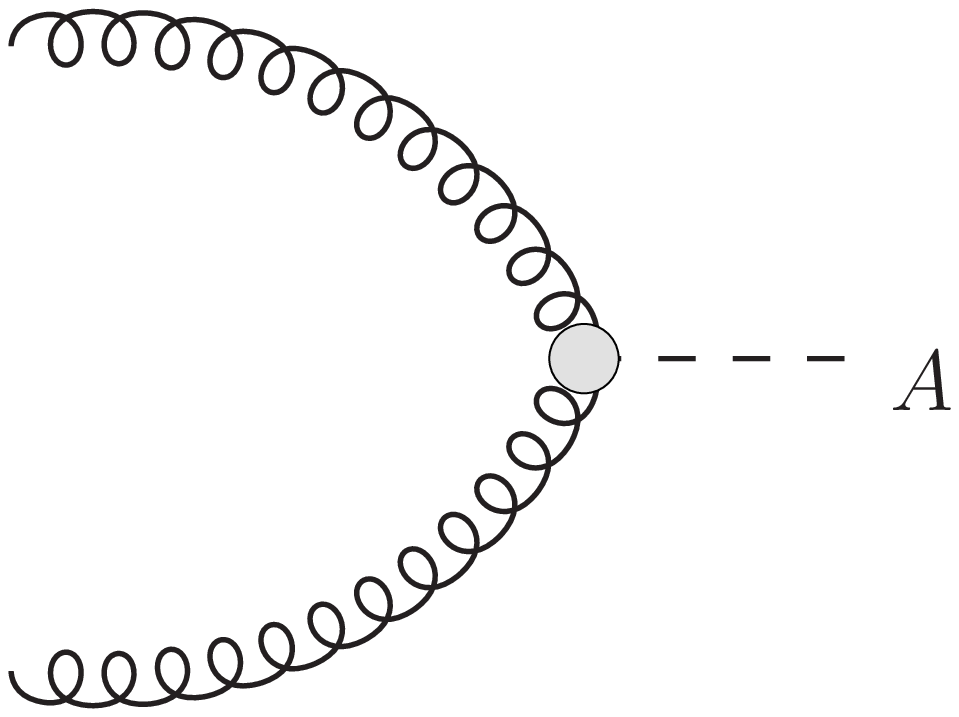} &
      \includegraphics[width=8em]{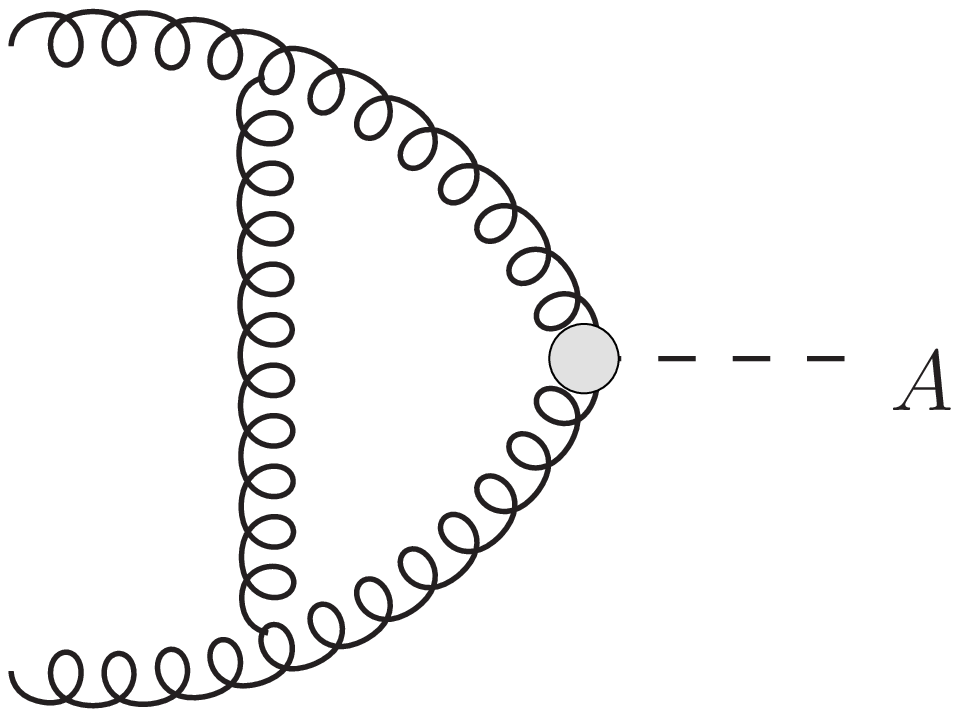} &
      \includegraphics[width=8em]{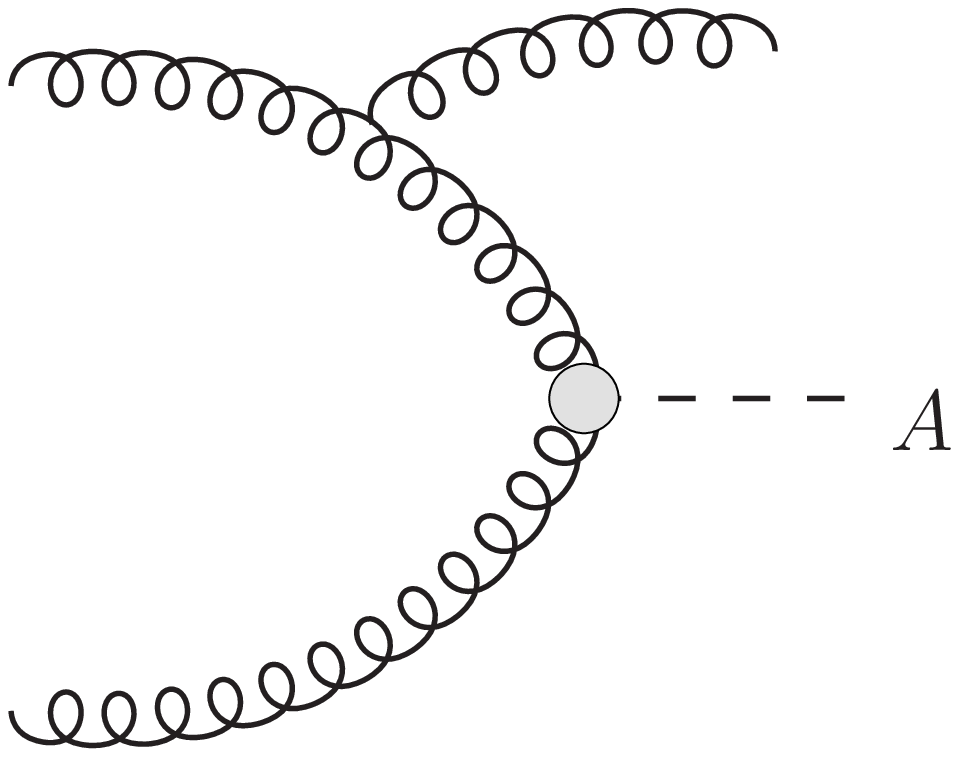} \\
      (a) & (b) & (c)\\[1em]
      \end{tabular}
      \\
      \begin{tabular}{ccc}
      \includegraphics[width=8em]{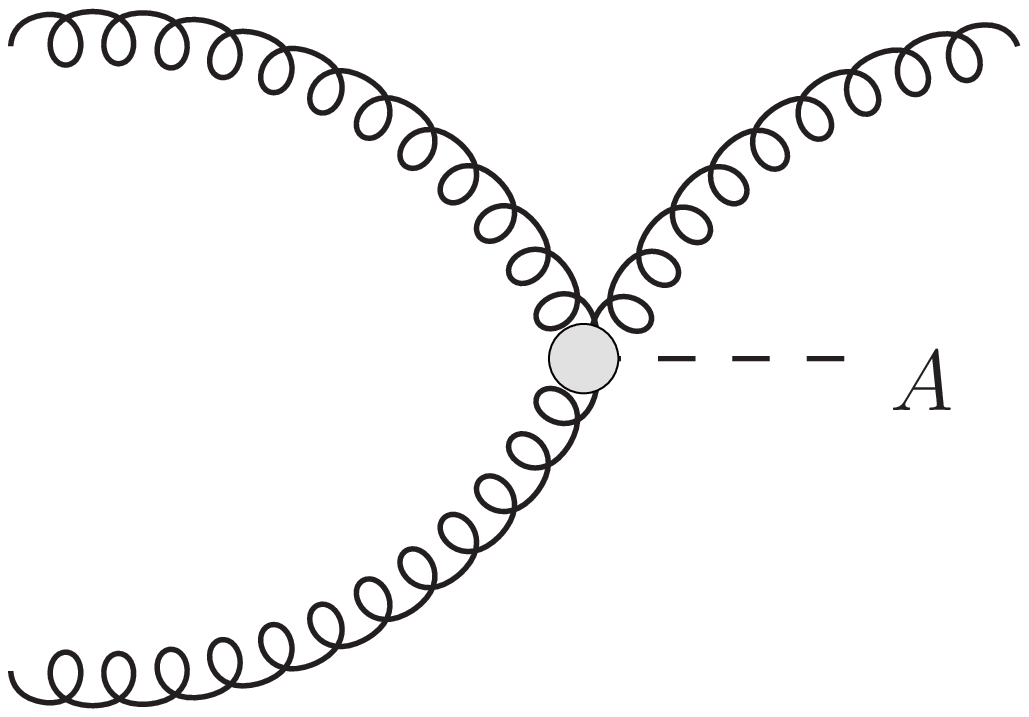} &
      \includegraphics[width=8em]{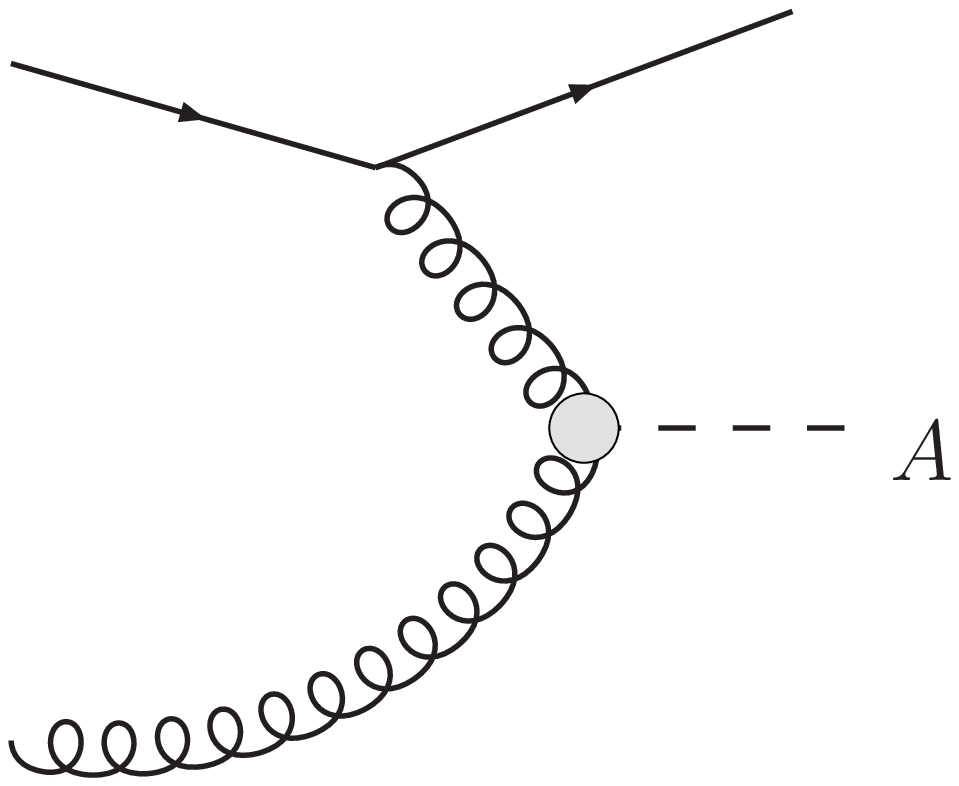} &
      \raisebox{.5em}{\includegraphics[width=9em]{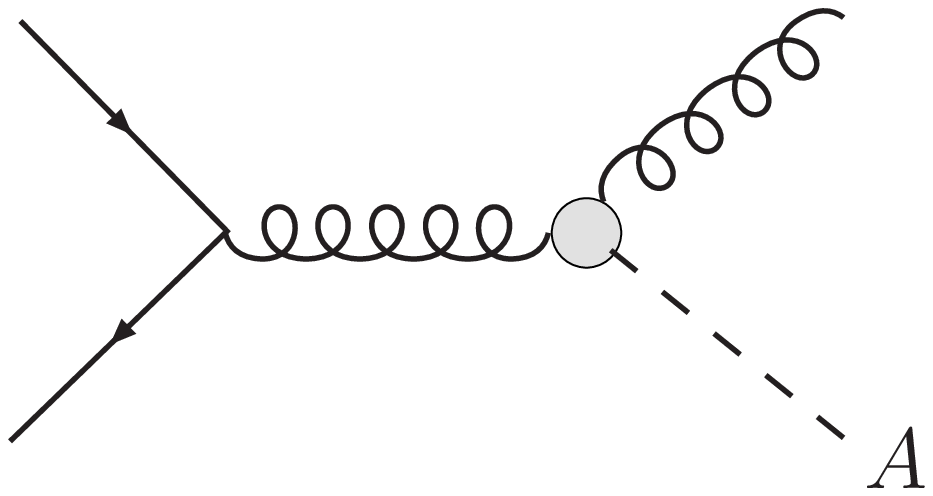}} \\
      (d) & (e) & (f)
      \end{tabular}
    \end{tabular}
    \caption[]{\sloppy
      Diagrams contributing to the gluon fusion process in the effective
      theory at \nlo{}.
      \label{fig::process}
    }
}

\subsection{SUSY contributions}\label{sec::susy}
We construct again an effective theory, this time by integrating out not
only the top quark, but also all the \susy{} particles. Since
the remaining degrees of freedom are the same as in the Standard Model
case of \sct{sec::eff}, the effective Lagrangian has exactly the same
form as in \eqn{eq::leffsm}, only with $\tilde C_1^{\rm SM}$ and $\tilde
C_2^{\rm SM}$ replaced by $\tilde C_1$ and $\tilde C_2$.

As in the \sm{} case, $\tilde C_2$ is zero through \nlo{}. It remains to
determine $\tilde C_1$ within the framework of \eqn{eq::lag} through
\nlo{} in $\alpha_s$.  The calculation of the \sm{} case in
Ref.\,\cite{Chetyrkin:1998mw} was done using Dimensional Regularization
as it is most convenient for loop calculations in {\abbrev
QCD}.  It is known that \dreg{} breaks \susy{} and requires
\susy{}-restoring finite counter terms in general. In the current case,
however, these are not required due to the absence of \susy{}
contributions at \lo{}.

On the other hand, one may apply Dimensional
Reduction~\cite{Siegel:1979wq} in order to regularize the divergent
integrals.  This is done by setting $D=4$ after contracting all (Lorentz
and spinor) indices~\cite{Jack:1994bn,Jack:1997sr}, while the loop
integrals are subsequently evaluated in $D=4-2\ep$ dimensions.

A particularly subtle issue is the treatment of $\gamma_5$ which occurs
in the Higgs--quark--anti-quark and the gluino--quark--squark vertex. We
anticommute these $\gamma_5$ matrices and use $\gamma_5^2=1$ until only
one of them remains in the Fermion trace. Working in \dreg{}, we may
follow the calculation of Ref.\,\cite{Chetyrkin:1998mw} which adopted
the method of Ref.\,\cite{'tHooft:fi,Akyeampong:1973xi,Larin:1993tq} for
the treatment of $\gamma_5$. This involves a finite counter term $Z_5^p$
in order to restore gauge invariance~\cite{Trueman:1979en}. Since there
are no \susy{} diagrams at \lo{}, $Z_5^p$ is given by the well-known \sm{}
expression quoted in Ref.\,\cite{Chetyrkin:1998mw}.

For the \dred{} calculation, on the other hand, we simply
set~\cite{'tHooft:fi,Larin:1993tq,Chetyrkin:1998mw}
\begin{equation}
\begin{split}
\gamma_5 = \frac{i}{4!}\varepsilon_{\mu\nu\rho\sigma}
\gamma^{\mu}
\gamma^{\nu}
\gamma^{\rho}
\gamma^{\sigma}\,.
\end{split}
\end{equation}
It has been shown that the occurrence of the Levi-Civita symbol
$\varepsilon_{\mu\nu\rho\sigma}$ can lead to inconsistencies when
implemented in \dred{}~\cite{Siegel:1980qs}.  We circumvent them by
keeping the genuinely 4-dimensional Levi-Civita symbol
$\vep_{\mu\nu\rho\sigma}$ uncontracted until after the renormalization
procedure, in close analogy to
Refs.\,\cite{Larin:1993tq,Chetyrkin:1998mw}. As opposed to the
calculation in \dreg{}~\cite{Chetyrkin:1998mw}, however, working in
\dred{} by definition does not involve terms of $\order{\ep=2-D/2}$ that
require the above-mentioned finite counter terms in order to restore
gauge invariance. We have verified that the scalar and pseudo-scalar
one-loop quark vertices are identical in the chiral limit.
Thus, the formulae of Ref.\,\cite{Chetyrkin:1998mw} for
projecting onto the coefficient function $\tilde C_1$ can be translated
to the \dred{} case by setting $D=4$ and ignoring the finite
renormalization of the pseudo-scalar current (i.e., $Z_5^p=1$).

We will explicitely demonstrate that the \sm{} result is recovered
(through $\order{\alpha_s^3}$) by making all the superpartner masses
infinitely heavy. It turns out that the {\abbrev SUSY} diagrams lead to
terms that do not vanish as $M_{\rm SUSY}\to \infty$. However, they
cancel with the counter terms of the {\abbrev SM} diagrams to give the
correct {\abbrev SM} result.

The generalization of this approach for the implementation
of $\gamma_5$ within \dred{} through higher orders is very desirable, of
course, because it greatly simplifies precision calculations in \susy{}
models as they might be required by future experimental data.

\section{Results}\label{sec::results}
The evaluation of the diagrams proceeds in complete analogy to
Ref.\,\cite{Harlander:2004tp}. At \nlo{}, one needs to evaluate massive
2-loop diagrams with vanishing external momenta. This is possible in a
fully analytic way with the help of the algorithm of
Ref.\,\cite{Davydychev:1992mt}\footnote{We are indebted to
M.~Steinhauser for providing us with his implementation of this
algorithm in the framework of {\tt MATAD}~\cite{Steinhauser:2000ry}.}. As
a cross check, we also calculated the diagrams in the limit $m_t\ll
\mstop{1}\ll \mstop{2}\ll \mgluino{}$ by using automated asymptotic
expansions~\cite{exp} and found agreement with the corresponding
expansion of the analytical result.

Subsequently, the bare coupling constant within the \susy{}-\qcd{}
theory is transformed to its renormalized 5-flavor \qcd{} expression in
the \msbar{} scheme as described in Ref.\,\cite{Harlander:2004tp}.

In \dred{}, and using the prescription for the treatment of $\gamma_5$
described above, we find the following contribution of all \sm{}
diagrams to the coefficient function:
\begin{equation}
\begin{split}
   \tilde C_1\big|_{\mbox{\scriptsize SM-dias}} &= -
   \frac{\alpha_s}{16\pi}\cot\beta\bigg\{
   1 - \frac{2}{3} \api \bigg\} + \order{\alpha_s^3}\,.
   \label{eq::smdias}
\end{split}
\end{equation}
The second term in the curly brackets arises from the \mssm{} expression
for the quark mass counter term. Using the \sm{} expression instead, we
recover the well-known result of Eq.~(\ref{eq::c1sm}).
The \susy{} diagrams add up to
\begin{equation}
\begin{split}
   \tilde C_1\big|_{\mbox{\scriptsize SUSY-dias}} &= -
   \frac{\alpha_s}{16\pi}\cot\beta\bigg\{ \frac{2}{3} \api \bigg\} +
   \order{\frac{m_t^2}{M^2}} + \order{\alpha_s^3}\,,\qquad
   M\in\{\mstop{},\mgluino\}\,,
   \label{eq::susydias}
\end{split}
\end{equation}
such that indeed the \sm{} limit is recovered as the \susy{} masses are
decoupled:
\begin{equation}
\begin{split}
\tilde C_1^{\rm SM} &= \lim_{M_{\rm SUSY}\to\infty}\left( \tilde
C_1\big|_{\mbox{\scriptsize SM-dias}} + \tilde C_1\big|_{
\mbox{\scriptsize SUSY-dias}}\right)\,.
\end{split}
\end{equation}

The general \nlo{} result for $\tilde C_1$ can then be cast into the
following form:
\begin{equation}
\begin{split}
\tilde C_1 &= - \frac{\alpha_s}{16\pi}\cot\beta\bigg\{ 1 +
\api\,\frac{\muSUSY}{\mgluino}\left(\cot\beta + \tan\beta\right)\,
f(m_t,\mstop{1},\mstop{2},\mgluino)\bigg\} + \order{\alpha_s^3}\,.
\label{eq::c1res}
\end{split}
\end{equation}
A few observations may be worth pointing out:
\begin{itemize}
\item \eqn{eq::c1res} immediately shows that the \nlo{} corrections to
  the coefficient function will get more important w.r.t.\ the \lo{}
  term for large values of $\tan\beta$ and $\muSUSY$. On the other hand,
  large values of $\tan\beta$ increase the importance of bottom
  contributions even more, so that they usually obscure the squark
  effects.
\item The $\order{\alpha_s^2}$ corrections are proportional to
  $\muSUSY$, while the squark mixing angle $\theta_t$ (or, equivalently,
  the trilinear coupling $A_t$) drops out in the final result.  This is
  due to the axial U(1) Peccei-Quinn symmetry of the \susy{}
  potential~\cite{reuter,hunter} whose explicit breaking by the
  $\muSUSY$ term violates the Adler-Bardeen theorem. It provides another
  consistency check of our calculation.
\item As a result of the absence of mass terms at \lo{}, the explicit form of
  $f$ is independent of the mass renormalization scheme.
\item Eqs.~(\ref{eq::smdias}) and (\ref{eq::susydias}) are unchanged in
\dreg{} if the \sm{} diagrams are multiplied by the finite
renormalization constant $Z_5^p$ (see Sect.~\ref{sec::susy} and
Ref.\,\cite{Chetyrkin:1998mw}).
\end{itemize}
The general result for $f(m_t,\mstop{1},\mstop{2},\mgluino)$ is too
voluminous to be displayed here. Instead, we implemented $\tilde C_1$ in
the program {\tt
evalcsusy.f}~\cite{Harlander:2004tp}\footnote{\label{foot::evalcsusy} The
program is available from {\tt
http://www-ttp.physik.uni-karlsruhe.de/Progdata/ttp04/ttp04-19.}} and
analyze its behavior numerically as shown in \sct{sec::discussion}.

For practical purposes, it might be useful to provide the analytical
expression of the function $f$ for some limiting
cases. Defining two mass scales $m\ll M$, and
\newcommand{\lx}{l_{x}}
\begin{equation}
\begin{split}
x = \frac{m^2}{M^2}\,,\qquad 
\lx = \ln(x)\,,
\end{split}
\end{equation}
we find
\begin{align}
    &&& f(m,m,m,m) &=&&&  \frac{5}{18} - \frac{9}{2}\,S_2 =
    -0.894176\ldots\,,\\
    &&& f(m,M,M,M) &=&&& 
    - \frac{1}{3}
    - x\left( \frac{35}{54} - \frac{5}{36}\,\lx \right)
    - x^2\left( \frac{41}{1800} + \frac{7}{60}\,\lx \right)
    + \ldots\,,\\
    &&& f(m,m,m,M) &=&&&  \frac{2}{3} + \frac{2}{3}\,\lx
    + x \left( \frac{7}{6} -
    \frac{1}{3}\,\zeta_2 + \frac{13}{6}\,\lx - \frac{1}{6}\,\lx^2 \right)
    + \ldots\,,\\
    &&& f(M,m,m,m) &=&&&   \frac{3}{2} + \frac{3}{2}\,\lx 
    + x\left( \frac{11}{3} + 3\,\zeta_2 + \frac{29}{3}\,\lx +
    \frac{3}{2}\,\lx^2 \right) + \ldots\,,\\
    &&& f(m,M,M,m) &=&&&  - \frac{13}{6}\,x
    - x^2 \left(\frac{22}{3} + 3\,\zeta_2 + \frac{31}{3}\lx +
    \frac{3}{2}\,\lx^2 \right) + \ldots\,,\\
    &&& f(m,m,M,M) &=&&& - \frac{2}{3}
      - x\left(\frac{181}{108} + \frac{13}{36}\,\lx\right) 
+ \ldots\,,
\end{align}
where the dots denote higher orders in $x$, and
\begin{equation}
\begin{split}
S_2 = \frac{4}{9\sqrt{3}}\,{\rm Cl}_2\left(\frac{\pi}{3}\right)\,,\qquad
\zeta_2 = \frac{\pi^2}{6}\,.
\end{split}
\end{equation}

Insertion of $f(m,m,m,M)$ into \eqn{eq::c1res} shows that, in contrast
to scalar Higgs
production~\cite{Harlander:2003bb,Harlander:2003kf,Harlander:2004tp},
the result for $\tilde C_1$ is well-behaved as $\mgluino\to \infty$ at
finite $m$. This is because there are no squark contributions at \lo{}
which could affect the renormalization of the \nlo{} terms.  The same
holds for $\mstop{}\to\infty$ as can be seen from the form of
$f(m,M,M,m)$. Note, however, that the result is logarithmically
divergent in the case where $m_t\to \infty$ with the other
masses fixed, see $f(M,m,m,m)$. This is due to the linear mass
dependence of the top Yukawa coupling and can be observed already in the
pure \sm{} contributions at higher orders in $\alpha_s$.

\section{Discussion}\label{sec::discussion}

\subsection[Coefficient function $\tilde C_1$]{%
  Coefficient function \bld{\tilde C_1}}

We will now discuss the numerical effect of the newly evaluated terms.
First, we investigate the dependence of $\tilde C_1$ on the squark and
gluino masses. To this aim, it is convenient to consider
the leading and the \nlo{} term separately:
\begin{equation}
\begin{split}
\tilde C_1 &= -\frac{\alpha_s}{16\pi}\left( \tilde c_1^{(0)} +
\frac{\alpha_s}{\pi} \,\tilde c_1^{(1)} \right) + \order{\alpha_s^3}\,.
\label{eq::c1pert}
\end{split}
\end{equation}
According to \eqn{eq::c1res},
\begin{equation}
\begin{split}
\tilde c_1^{(0)} &= \cot\beta\,.
\end{split}
\end{equation}

\fig{fig::c1mgluino} shows $\tilde c_1^{(1)}$ as a function of
$\mgluino$ for various values of $\mstop{1}$ and $\mstop{2}$. In
\fig{fig::c1mstop2}, on the other hand, we consider it as a function of
one of the stop masses ($\mstop{A}$) while fixing $\mgluino{}$ and the
other stop mass ($\mstop{B}$) at a few representative values; $A$ and
$B$ assume the values $1$ and $2$, depending on whether $\mstop{A}$ is
larger or smaller than $\mstop{B}$. Note that $\tilde c_1^{(1)}$ is
symmetric in $\mstop{1}$ and $\mstop{2}$ since it does not depend on the
squark mixing angle $\theta_t$ as mentioned above. Finally, in
\fig{fig::c1mstop}, $\mstop{1}=\mstop{2}=\mstop{}$ is varied for certain
choices of $\mgluino$.

The general structure is quite similar in all figures: $\tilde c_1^{(1)}$
is of the order of $-0.5$ for moderate values of the \susy{} masses. As
they increase, $\tilde c_1^{(1)}$ tends to zero in agreement with the
general discussion above. In order to demonstrate this behavior more
clearly, \fig{fig::c1log} adopts the same parameters as
\fig{fig::c1mgluino}, but extends up to very large values of the gluino
mass.

\FIGURE{%
  \psfrag{C}{ $\tilde{c}_1^{(1)}$}
  \psfrag{Mgluino}{ $m_{\tilde{g}}$/GeV}
  \includegraphics[width=0.7\textwidth]{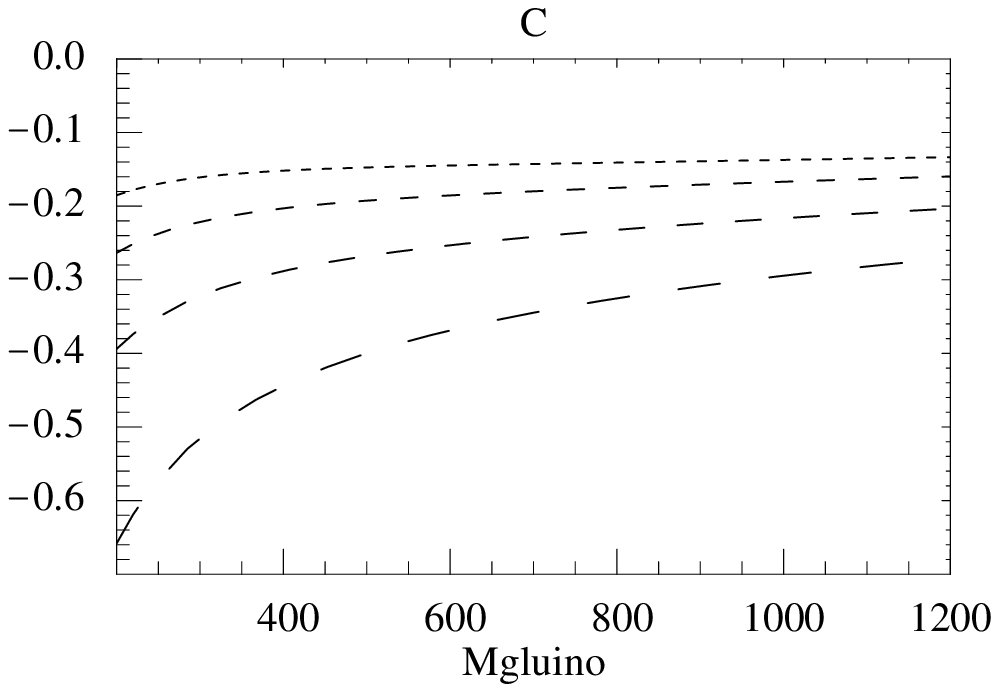}
\caption[]{\label{fig::c1mgluino}$\tilde c_1^{(1)}$ as a function of the
  gluino mass for $m_t=173$\,GeV, $\muSUSY = 150$\,GeV, $\tan\beta=3$,
  and $(\mstop{1},\mstop{2}) =
  (200,200)/(200,400)/(200,600)/(400,600)$\,GeV [long/\ldots/short
  dashes].}  }

\FIGURE{%
  \psfrag{C}{ $\tilde{c}_1^{(1)}$}
  \psfrag{Mstop2}{ $\mstop{A}$/GeV}
  \includegraphics[width=0.7\textwidth]{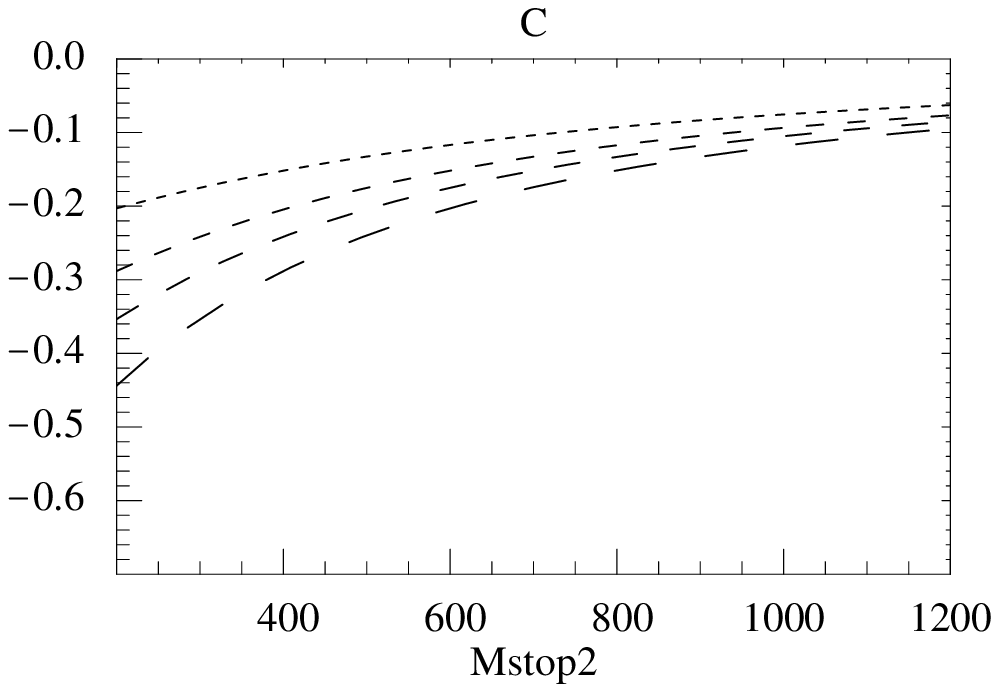}
\caption[]{\label{fig::c1mstop2}$\tilde c_1^{(1)}$ as a function of
  $\mstop{A}$ for $m_t=173$\,GeV, $\muSUSY = 150$\,GeV, $\tan\beta=3$,
  $\mgluino = 400$\,GeV, and $\mstop{B} = 200/300/400/600$\,GeV
  [long/\ldots/short dashes]; [$(A,B)=(1,2)$ if $\mstop{A}\leq\mstop{B}$,
  otherwise $(A,B)=(2,1)$].}  }

\FIGURE{%
  \psfrag{C}{ $\tilde{c}_1^{(1)}$}
  \psfrag{M12}{ $\mstop{}$/GeV}
  \includegraphics[width=0.7\textwidth]{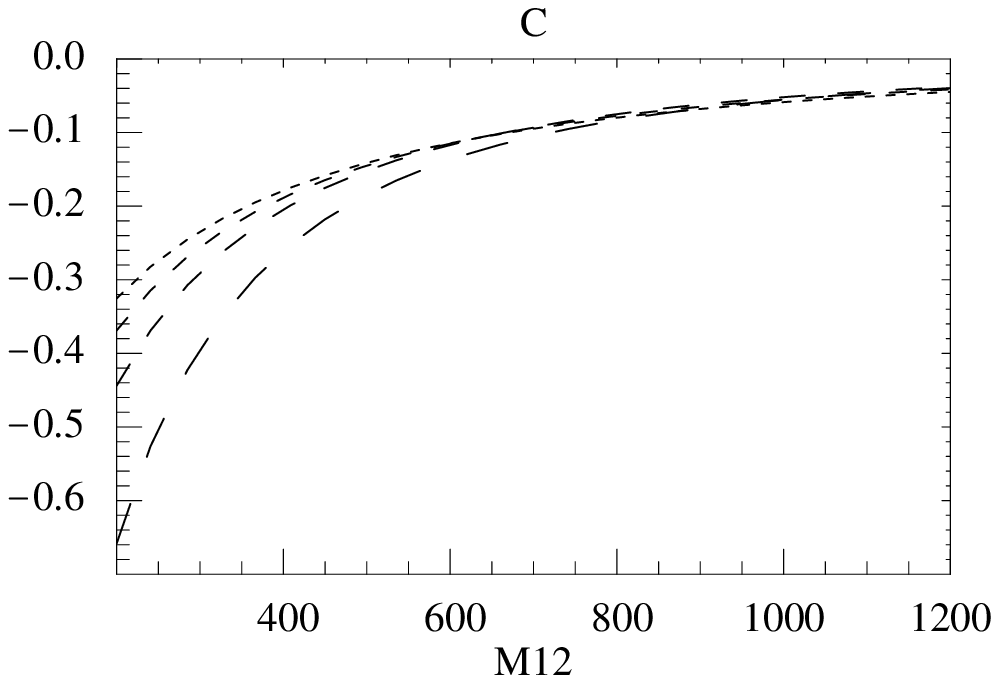}
\caption[]{\label{fig::c1mstop}$\tilde c_1^{(1)}$ as a function of
  $\mstop{}\equiv \mstop{1}=\mstop{2}$ for $m_t=173$\,GeV, $\muSUSY =
  150$\,GeV, $\tan\beta=3$, $\mgluino = 200/400/600/800$\,GeV
  [long/\ldots/short dashes].}  }

\FIGURE{%
  \psfrag{C}{ $\tilde{c}_1^{(1)}$}
  \psfrag{Mgluino}{ $m_{\tilde{g}}$/GeV}
  \includegraphics[width=0.7\textwidth]{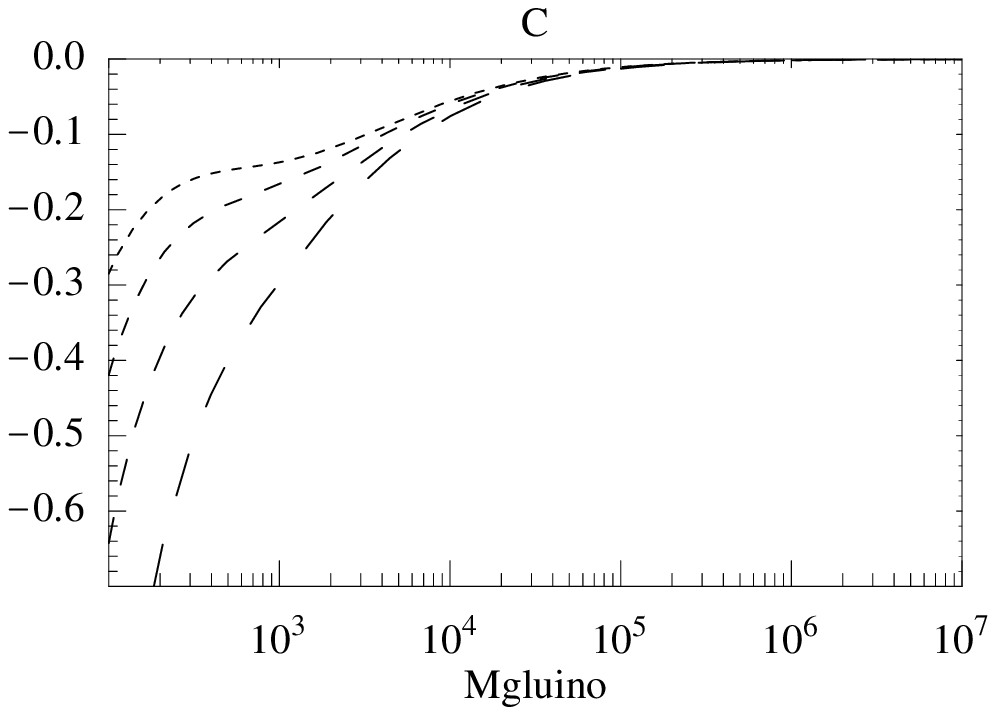}
\caption[]{\label{fig::c1log}Same as \fig{fig::c1mgluino}, but extended
  abscissa in order to demonstrate the asymptotic
  behavior of $\tilde c_1^{(1)}$.}  }

For the discussion of the numerical effects on the cross section in the
following section, we adopt a scenario similar to the $m_h^{\rm max}$
benchmark scenario~\cite{Carena:1999xa}, corresponding to
\begin{equation}
\begin{split}
\mstop{1}=826\,\mbox{GeV}\,,\qquad \mstop{2}=1172\,\mbox{GeV}\,,\qquad
\mgluino=800\,\mbox{GeV}\,.
\label{eq::mhmax}
\end{split}
\end{equation}
In the original definition of this scenario, $\muSUSY$ was set to
$-200$\,GeV; here, however, we will allow for a variation of this
parameter between $\pm 1$\,TeV.  In addition, we will assume
$\tan\beta=3$; the top mass is set to $m_t=173$\,GeV.\footnote{We remark
that, for $M_A\lesssim 200$\,GeV, this choice of parameters leads to a
conflict between the theoretical value of the light Higgs mass $m_h$
(evaluated with {\tt FeynHiggs}~\cite{Heinemeyer:1998yj}, for example)
and its current experimental lower limit; nevertheless, for the sake of
generality, we will vary $M_A$ between $100$ and $300$\,GeV in our
numerical analyses below. }

 The 2-loop term of
the coefficient function is then given by
\begin{equation}
\begin{split}
\tilde c_1^{(1)} &= -0.392\,\frac{\muSUSY}{1\mbox{TeV}}\,.
\label{eq::c1mhmax}
\end{split}
\end{equation}

\subsection{Effects on the cross section}
The inclusive \nlo{} hadronic cross section for pseudo-scalar Higgs
production receives contributions from the subprocesses $gg\to A(+g)$,
$qg\to Aq$, and $q\bar q\to Ag$.  At this order, squarks
only affect the process $gg\to A$, because of the antisymmetric structure
of the ${\tilde q}_i{\tilde q}_jA$ coupling, cf.\
Eqs.\,(\ref{eq::lags}), (\ref{eq::couplings}).
Recalling the notation of \eqn{eq::hadpart}, we write
\begin{equation}
\begin{split}
\hat\sigma_{gg}(x)
 &= \hat\sigma_{tb}(x)
+ \Delta\hat\sigma_{\tilde tt}(x)
+ \Delta\hat\sigma_{\tilde tb}(x)
+ \order{\alpha_s^4}\,,
\end{split}
\end{equation}
where $\hat\sigma_{tb}$ denotes the contributions arising from the top and
bottom mediated gluon-Higgs couplings. It can be evaluated through
\nlo{} for arbitrary top, bottom, and Higgs masses with the help of the
{\tt FORTRAN} program {\tt HIGLU}. The same is true for the $qg$ and the
$q\bar q$
sub-processes (see Figs.\,\ref{fig::lonlo}\,(e) and (f)).

$\Delta\hat\sigma_{\tilde tt}$ and $\Delta\hat\sigma_{\tilde tb}$ are the
effects arising from the interference of stop-induced with top- and
bottom-induced amplitudes:
\begin{equation}
\begin{split}
\Delta\hat\sigma_{\tilde tq} \sim \Re\left({\cal M}_{\tilde t}^{(1)\ast}
{\cal M}_q^{(0)}\right)\,,\qquad q\in\{b,t\}\,.
\end{split}
\end{equation}
${\cal M}_q^{(0)}$ is expressed in terms of ${\cal A}_q$ from
\eqn{eq::LO}, while for ${\cal M}_{\tilde t}^{(1)}$, we use the
expression evaluated in the effective theory to obtain
\begin{equation}
\begin{split}
\Delta\hat\sigma_{\tilde tq}(x) &=\frac{\pi}{128v^2}
 \left(\api\right)^3\Re\left(\tilde c_1^{(1)}\,{\cal
 A}_q(\tau_q)\right)\,
 \delta(1-x)\,.
\end{split}
\end{equation}
Note that ${\cal M}_{\tilde t}^{(1)}$ has a branch cut at
$M_A=2m_t$. Thus, we expect our result to be valid for $M_A<2m_t$.
Recall, however, that in the \sm{} case, the heavy top limit still
provides an excellent approximation for Higgs masses much larger than
$2m_t$~\cite{Spira:1995rr,Spira:1997dg,Kramer:1996iq,Harlander:2003xy}.

To study the numerical effects, we consider the modified $m_h^{\rm max}$
scenario quoted in \eqn{eq::mhmax}. \fig{fig::rat1} shows the relative
size of the top squark effects to the total \nlo{} cross section. We
note that even for $|\muSUSY|=1$\,TeV, they hardly exceed 4\%.

\fig{fig::splittb} shows separately the effects of the top-stop and the
bottom-stop interference terms, again relative to the total \nlo{} cross
section, for $\mu=1$\,TeV. They are of similar order in magnitude, but
opposite in sign, thus cancelling each other to a certain extent.

Finally, \fig{fig::mhmaxsig} shows the inclusive cross section through
\nlo{} for the modified $m_h^{\rm max}$ scenario defined in and below
\eqn{eq::mhmax}, including effects of top and bottom quarks (solid line)
as well as top squarks (dashed lines) for $\muSUSY=\pm 1$\,TeV.

\FIGURE{%
    \begin{tabular}{c}
      \includegraphics[bb=75 300 365 535,width=.75\textwidth]{%
	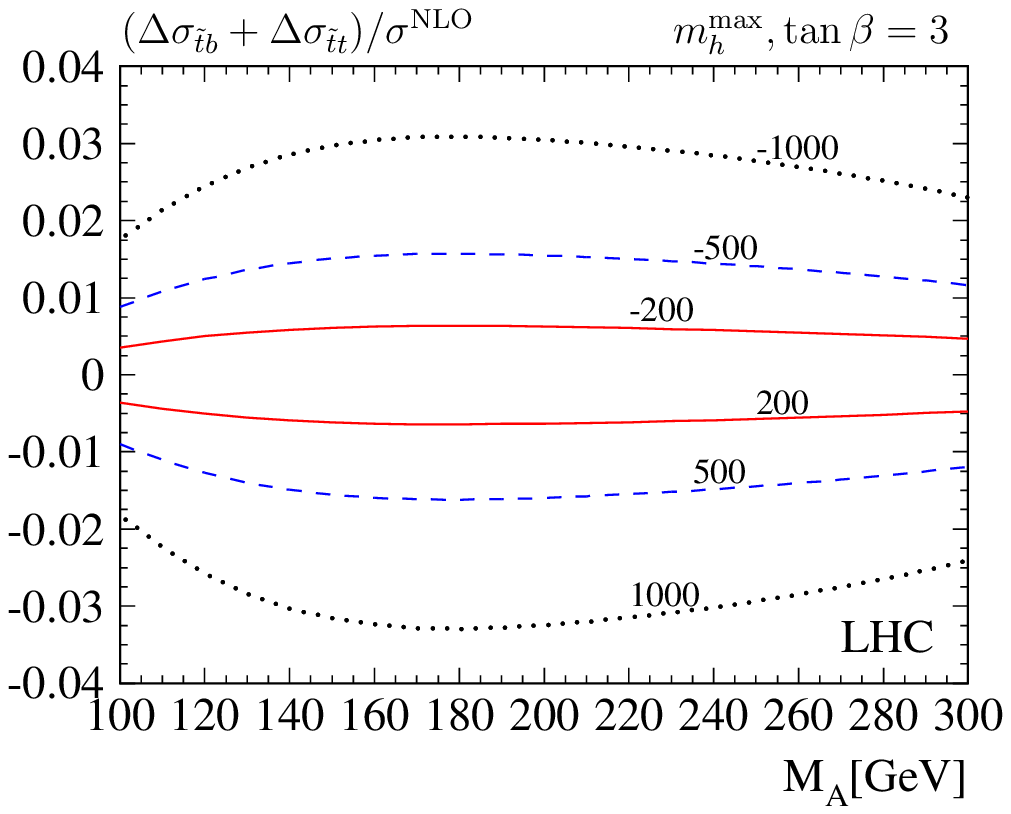}
    \end{tabular}
    \parbox{.9\textwidth}{
      \caption[]{\label{fig::rat1}\sloppy Effects of the
	top-stop and bottom-stop interference terms relative to the
	total \nlo{} cross section for the scenario defined in and below
	\eqn{eq::mhmax}. The numbers above the graphs denote the value
	of $\muSUSY$ in GeV.}
}}

\FIGURE{%
    \begin{tabular}{c}
      \includegraphics[bb=75 300 365 535,width=.7\textwidth]{%
	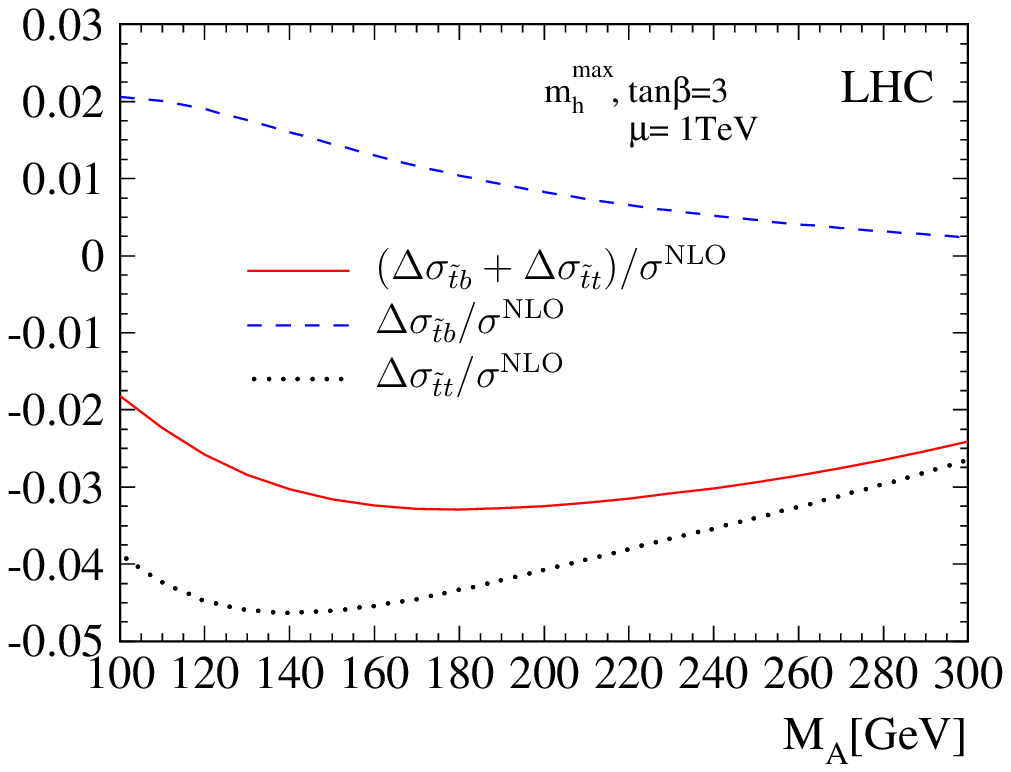}
    \end{tabular}
    \parbox{.9\textwidth}{
      \caption[]{\label{fig::splittb}\sloppy
	Relative effect of the top-stop (dotted) and bottom-stop (dashed)
	interference terms, as well as their sum (solid), for the
	scenario defined in and below \eqn{eq::mhmax}.
        }}
}

\FIGURE{%
    \begin{tabular}{c}
      \includegraphics[bb=110 240 490 560,width=.75\textwidth]{%
	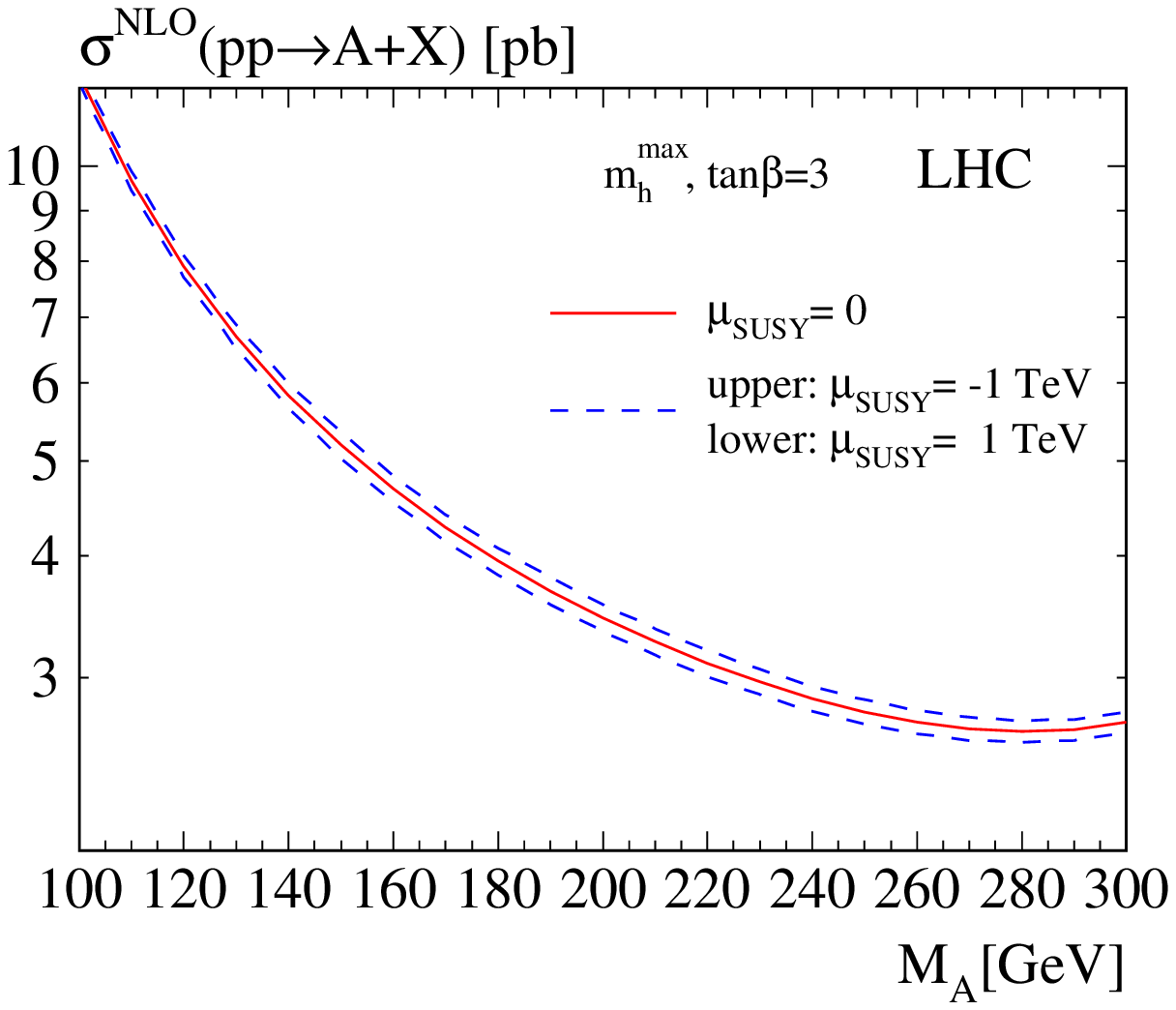}
    \end{tabular}
    \parbox{.9\textwidth}{
      \caption[]{\label{fig::mhmaxsig}\sloppy Total cross section at
	\nlo{} including top and bottom effects (solid). Stop effects
	(dashed) are shown for the scenario defined in and below
	\eqn{eq::mhmax} for two choices of $\muSUSY$. }}
}

\section{Conclusions}\label{sec::conclusions}
The corrections to the effective Higgs-gluon coupling for a
pseudo-scalar Higgs boson have been evaluated in the \mssm{} through
first order in the strong coupling constant $\alpha_s$, taking into
account effects of top and bottom quarks as well as top squarks. The
\nlo{} corrections are proportional to $\muSUSY$ as expected from the
symmetries of the \susy{} potential.  The numerical effects were studied
within a specific \susy{} scenario, derived from the $m_h^{\rm max}$
scenario of Ref.\,\cite{Carena:1999xa}; further studies can be performed
easily using the publicly available numerical routine {\tt
evalcsusy.f}~(see footnote on page \pageref{foot::evalcsusy}).

The calculation also addresses a technical issue, since it involves the
$\gamma_5$ matrix in a non-trivial way. In analogy to
Ref.\,\cite{Chetyrkin:1998mw}, we avoided the contraction of the
Levi-Civita symbol with $D$ dimensional quantities. We argued that the
calculation in \dred{} does not require finite counter terms as opposed
to the \dreg{} approach, provided the underlying theory is
supersymmetric.  The result was confirmed by evaluating the diagrams in
\dreg{} and including the proper counter term for $\gamma_5$.

It would be interesting to investigate these observations in more
detail, in particular to prove the validity of our implementation of
$\gamma_5$ within \dred{} in a rigorous way. Further corroboration could
be obtained from its application at second order $\alpha_s$. This
corresponds to the evaluation of $\tilde C_1$ at three loops for which
the technical tools are in principle available~\cite{Harlander:1998dq}.

From the phenomenological point of view, inclusion not only of bottom
but also sbottom quarks would be desirable. This could be done in the
heavy sbottom limit using asymptotic expansions of Feynman diagrams.
Moreover, one could calculate the photonic decay rate of the {\abbrev
CP}-odd Higgs boson in a very similar fashion.  There, however, one does
not need to rely strictly on the effective Lagrangian, but could
evaluate the first few terms of a Taylor expansion in $M_A^2/M^2$ along
the lines of Ref.\,\cite{Steinhauser:1996wy}, where
$M\in\{m_t,\mstop{},\mgluino\}$.

\paragraph{Acknowledgments.}
We are grateful to K.~Chetyrkin for various helpful comments and advice.
We would like to thank S.~Heinemeyer for his advice on \susy{} benchmark
scenarios, M.~Steinhauser for enlightening discussions concerning the
treatment of $\gamma_5$, and D.~St\"ockinger for his constructive
comments on the manuscript. We are particularly indebted to J.~Reuter
for his help in interpreting the structure of \eqn{eq::c1res}, and to
M.~Spira for clarifications concerning the \dreg{} approach in this
calculation, and various suggestions for improvement.  Further thanks go
to A.~Djouadi and J.~K\"uhn for encouragement and inspiring
conversations.

We kindly acknowledge financial support by {\it Deutsche
  Forschungsgemeinschaft}
(contract HA\,2990/2-1, {\it Emmy Noether program}).

\def\app#1#2#3{{\it Act.~Phys.~Pol.~}\jref{\bf B #1}{#2}{#3}}
\def\apa#1#2#3{{\it Act.~Phys.~Austr.~}\jref{\bf#1}{#2}{#3}}
\def\annphys#1#2#3{{\it Ann.~Phys.~}\jref{\bf #1}{#2}{#3}}
\def\cmp#1#2#3{{\it Comm.~Math.~Phys.~}\jref{\bf #1}{#2}{#3}}
\def\cpc#1#2#3{{\it Comp.~Phys.~Commun.~}\jref{\bf #1}{#2}{#3}}
\def\epjc#1#2#3{{\it Eur.\ Phys.\ J.\ }\jref{\bf C #1}{#2}{#3}}
\def\fortp#1#2#3{{\it Fortschr.~Phys.~}\jref{\bf#1}{#2}{#3}}
\def\ijmpc#1#2#3{{\it Int.~J.~Mod.~Phys.~}\jref{\bf C #1}{#2}{#3}}
\def\ijmpa#1#2#3{{\it Int.~J.~Mod.~Phys.~}\jref{\bf A #1}{#2}{#3}}
\def\jcp#1#2#3{{\it J.~Comp.~Phys.~}\jref{\bf #1}{#2}{#3}}
\def\jetp#1#2#3{{\it JETP~Lett.~}\jref{\bf #1}{#2}{#3}}
\def\jhep#1#2#3{{\small\it JHEP~}\jref{\bf #1}{#2}{#3}}
\def\mpl#1#2#3{{\it Mod.~Phys.~Lett.~}\jref{\bf A #1}{#2}{#3}}
\def\nima#1#2#3{{\it Nucl.~Inst.~Meth.~}\jref{\bf A #1}{#2}{#3}}
\def\npb#1#2#3{{\it Nucl.~Phys.~}\jref{\bf B #1}{#2}{#3}}
\def\nca#1#2#3{{\it Nuovo~Cim.~}\jref{\bf #1A}{#2}{#3}}
\def\plb#1#2#3{{\it Phys.~Lett.~}\jref{\bf B #1}{#2}{#3}}
\def\prc#1#2#3{{\it Phys.~Reports }\jref{\bf #1}{#2}{#3}}
\def\prd#1#2#3{{\it Phys.~Rev.~}\jref{\bf D #1}{#2}{#3}}
\def\pR#1#2#3{{\it Phys.~Rev.~}\jref{\bf #1}{#2}{#3}}
\def\prl#1#2#3{{\it Phys.~Rev.~Lett.~}\jref{\bf #1}{#2}{#3}}
\def\pr#1#2#3{{\it Phys.~Reports }\jref{\bf #1}{#2}{#3}}
\def\ptp#1#2#3{{\it Prog.~Theor.~Phys.~}\jref{\bf #1}{#2}{#3}}
\def\ppnp#1#2#3{{\it Prog.~Part.~Nucl.~Phys.~}\jref{\bf #1}{#2}{#3}}
\def\rmp#1#2#3{{\it Rev.~Mod.~Phys.~}\jref{\bf #1}{#2}{#3}}
\def\sovnp#1#2#3{{\it Sov.~J.~Nucl.~Phys.~}\jref{\bf #1}{#2}{#3}}
\def\sovus#1#2#3{{\it Sov.~Phys.~Usp.~}\jref{\bf #1}{#2}{#3}}
\def\tmf#1#2#3{{\it Teor.~Mat.~Fiz.~}\jref{\bf #1}{#2}{#3}}
\def\tmp#1#2#3{{\it Theor.~Math.~Phys.~}\jref{\bf #1}{#2}{#3}}
\def\yadfiz#1#2#3{{\it Yad.~Fiz.~}\jref{\bf #1}{#2}{#3}}
\def\zpc#1#2#3{{\it Z.~Phys.~}\jref{\bf C #1}{#2}{#3}}
\def\ibid#1#2#3{{ibid.~}\jref{\bf #1}{#2}{#3}}

\newcommand{\jref}[3]{{\bf #1}, #3 (#2)}
\newcommand{\bibentry}[4]{#1, #3.}
\newcommand{\arxiv}[1]{{\tt arXiv:#1}}

\end{document}